\documentclass{ws-p8-50x6-00}
\usepackage{graphicx}

\begin{document}

\title{Direct Photon Production in 158~{\it A}~GeV
   $^{208}$Pb\/+\/$^{208}$Pb Collisions}

\author{T.~Peitzmann}
\address{University of M{\"u}nster, D-48149 M{\"u}nster,
  Germany} 
\author{for the WA98 Collaboration}

\maketitle

\abstracts{
  A measurement of direct photon production in
  $^{208}$Pb\/+\/$^{208}$Pb collisions at 158~{\it A}~GeV has been
  carried out in the CERN WA98 experiment.  The invariant yield 
  of direct photons in central collisions is extracted
  as a function of transverse momentum
  in the interval $0.5 < p_T < 4$ GeV/c.  A significant direct photon
signal, compared to statistical and systematical errors, is seen
at  $ p_T > 1.5$ GeV/c. The results constitute the
  first observation of direct photons in ultrarelativistic 
  heavy-ion collisions which could be significant for diagnosis
of quark gluon plasma formation.
}

\section{Introduction}
The observation of a new phase of strongly interacting matter, the 
quark gluon plasma (QGP), is one 
of the most important goals of current nuclear physics research. 
To study QGP formation photons (both real and virtual) were one of the 
earliest proposed 
signatures \cite{Fei76,Shu78}. They 
are likely to escape from the system directly after 
production without further interaction, unlike hadrons. 
Thus, photons carry information on their emitting sources from 
throughout the entire collision history, including the initial hot 
and dense phase.
Recently, it was shown by Aurenche et al.~\cite{Aur98} that 
photon production rates in the QGP when calculated up to two loop
diagrams, 
are considerably greater than the earlier lowest order
estimates\cite{Kap91}.  Following this result,  
Srivastava~\cite{Sri99} has shown that at
sufficiently high initial temperatures
the photon yield from quark matter may significantly exceed
the contribution from the hadronic matter to provide a direct probe of
the quark matter phase.

A large number of measurements of prompt photon production at 
high transverse momentum ($p_{T} > 3 \, \mathrm{GeV}/c$) exist
for proton-proton, proton-antiproton, and proton-nucleus collisions
(see e.g. \cite{VW}). 
First attempts to observe direct photon production 
in ultrarelativistic heavy-ion collisions with oxygen and sulphur
beams found 
no significant excess~\cite{Ake90,Alb91,Bau96,Alb96}.
The WA80 collaboration \cite{Alb96} provided the most interesting
result with a $p_{T}$ dependent upper limit on the direct
photon production in S+Au collisions at 200$A$GeV. 
In this paper we report on the first observation of direct 
photon production in ultrarelativistic heavy-ion collisions.

\section{Data Analysis}

The results are from the CERN experiment WA98 \cite{misc:wa98:proposal:91} 
which consists
of large acceptance photon and hadron spectrometers. 
Photons are measured with the WA98 lead-glass photon detector,
LEDA, which consisted of 10,080 individual modules with 
photomultiplier readout. The detector was located at a distance of 
21.5~m from the target and covered the pseudorapidity interval 
$2.35 < \eta < 2.95$ $(y_{cm}=2.9)$. The particle identification was 
supplemented by a charged particle veto detector in front of LEDA.

The results presented here were obtained from an analysis of the
data taken with Pb beams in 1995 and 1996.
The 20\% most peripheral and the 10\% most central reactions have been 
selected from the minimum bias cross section 
($\sigma_{min.bias} \approx 6300 \, \mathrm{mb}$) 
using the measured transverse energy $E_{T}$. 
In total, $\approx 6.7 \cdot 10^{6}$ central and $\approx 4.3 \cdot 
10^{6}$ peripheral reactions have been analyzed.

The extraction of direct photons in the high multiplicity environment 
of heavy-ion collisions must 
be performed on a statistical basis by comparison of the measured
inclusive photon
spectra to the background expected from hadronic decays. 
Neutral pions and $\eta$ mesons are reconstructed via their $\gamma\gamma$ 
decay branch. 
For a detailed description of the 
detectors and the analysis procedure see \cite{wa98photonslong}.

\begin{figure}[bt]
\begin{center}
  \includegraphics[scale=0.4]{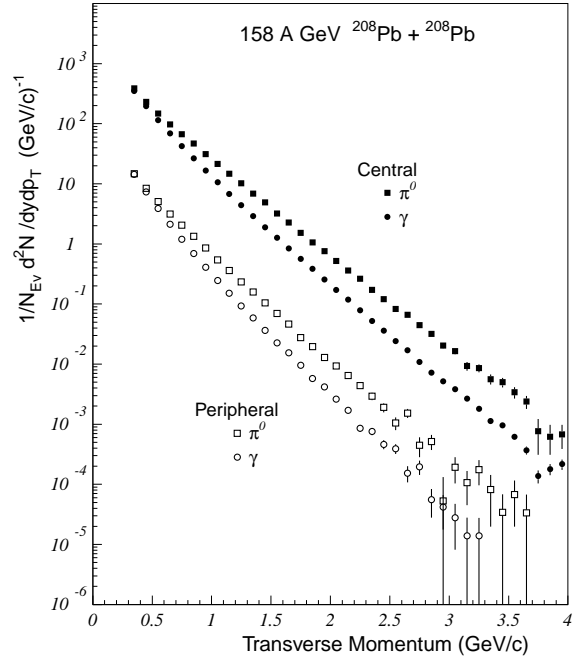}
\vspace{0.3cm}
  \caption{The inclusive photon (circles) and $\pi^0$ (squares) 
  transverse momentum distributions 
  for peripheral (open points) and central (solid points)
  158$~${\it A}~GeV $^{208}$Pb\/+\/$^{208}$Pb collisions. 
  The data have been corrected
  for efficiency and acceptance. Only statistical errors are shown.
  }
\label{fig:photon_pi0_pt}
\end{center}
\end{figure}

\begin{figure}[hbt]
\begin{center}
  \includegraphics[scale=0.4]{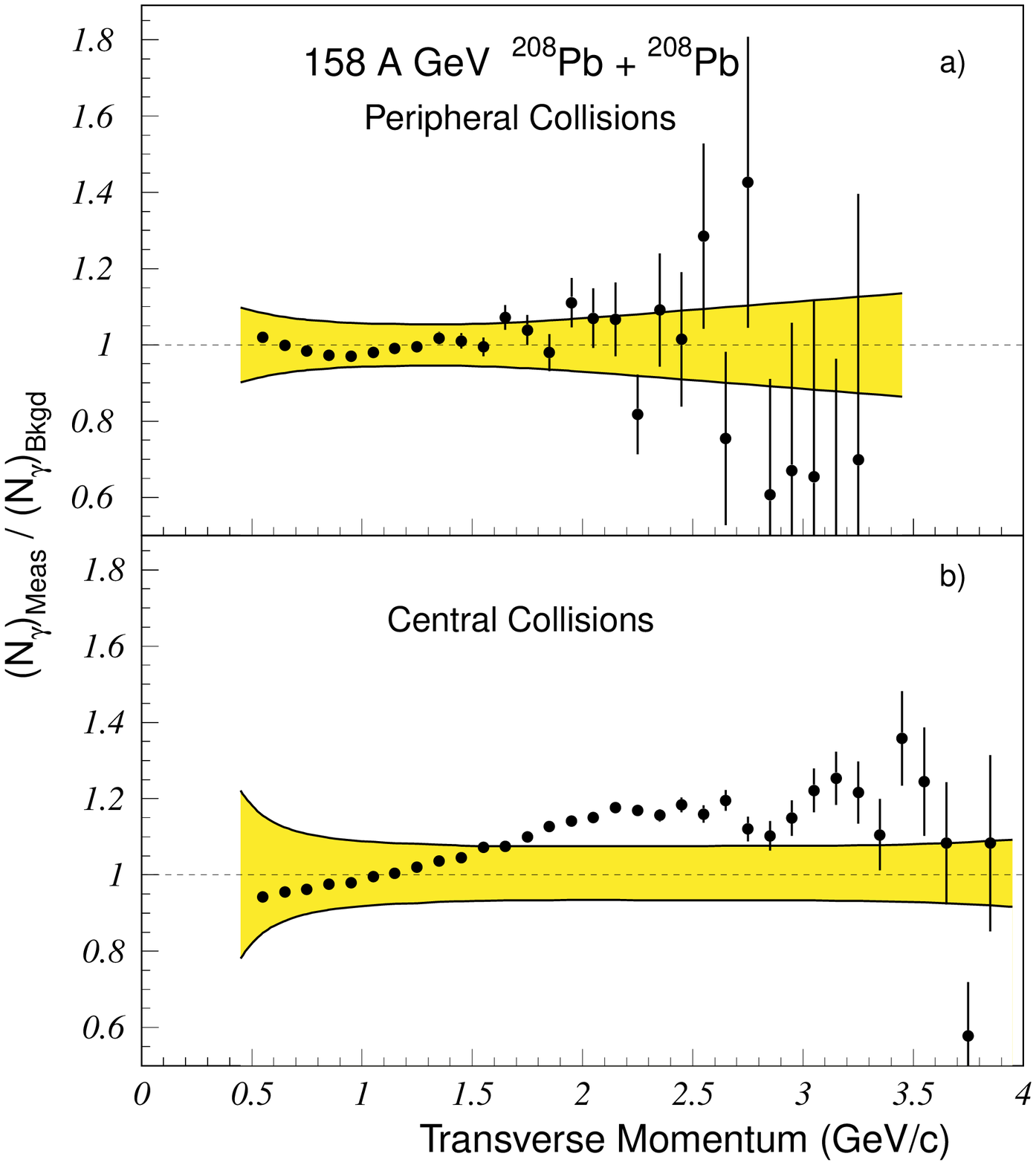}
\vspace{0.3cm}
\caption{The $\gamma_{\rm Meas}/\gamma_{\rm Bkgd}$ ratio 
  as a function of transverse momentum for peripheral (part a)) and
  central (part b)) 158~{\it A}~GeV $^{208}$Pb\/+\/$^{208}$Pb collisions. The
  errors on the data points indicate the statistical errors only. The
  $p_T$-dependent systematical errors are indicated by the shaded bands.
  }
\label{fig:gamma_excess}
\end{center}
\end{figure}

The final measured inclusive photon spectra are 
then compared to the calculated background photon 
spectra to check for a possible photon excess beyond that from 
long-lived radiative decays.
The background calculation is based on the 
measured  $\pi^0$ spectra and the measured $\eta/\pi^0$-ratio. 
The spectral shapes of other hadrons having radiative decays are calculated
assuming $m_{T}$-scaling~\cite{Bor76} with yields relative to
$\pi^0$'s taken from the literature.
It should be noted that the measured 
contribution (from $\pi^0$ and $\eta$) amounts to $\approx 97 \% $ of 
the total photon background. 

\section{Results}
Fig.~\ref{fig:photon_pi0_pt} shows the fully corrected inclusive photon 
spectra for peripheral and central collisions. The spectra cover the 
$p_{T}$ range of $0.3 - 4.0 \,\mathrm{GeV}/c$ (slightly less for 
peripheral collisions) and extend over six orders of magnitude. 
Fig.~\ref{fig:photon_pi0_pt} also shows the distributions of neutral 
pions which extend over a similar momentum range with slightly 
larger statistical errors. 

\begin{figure}[hbt]
\begin{center}
  \includegraphics[scale=0.4]{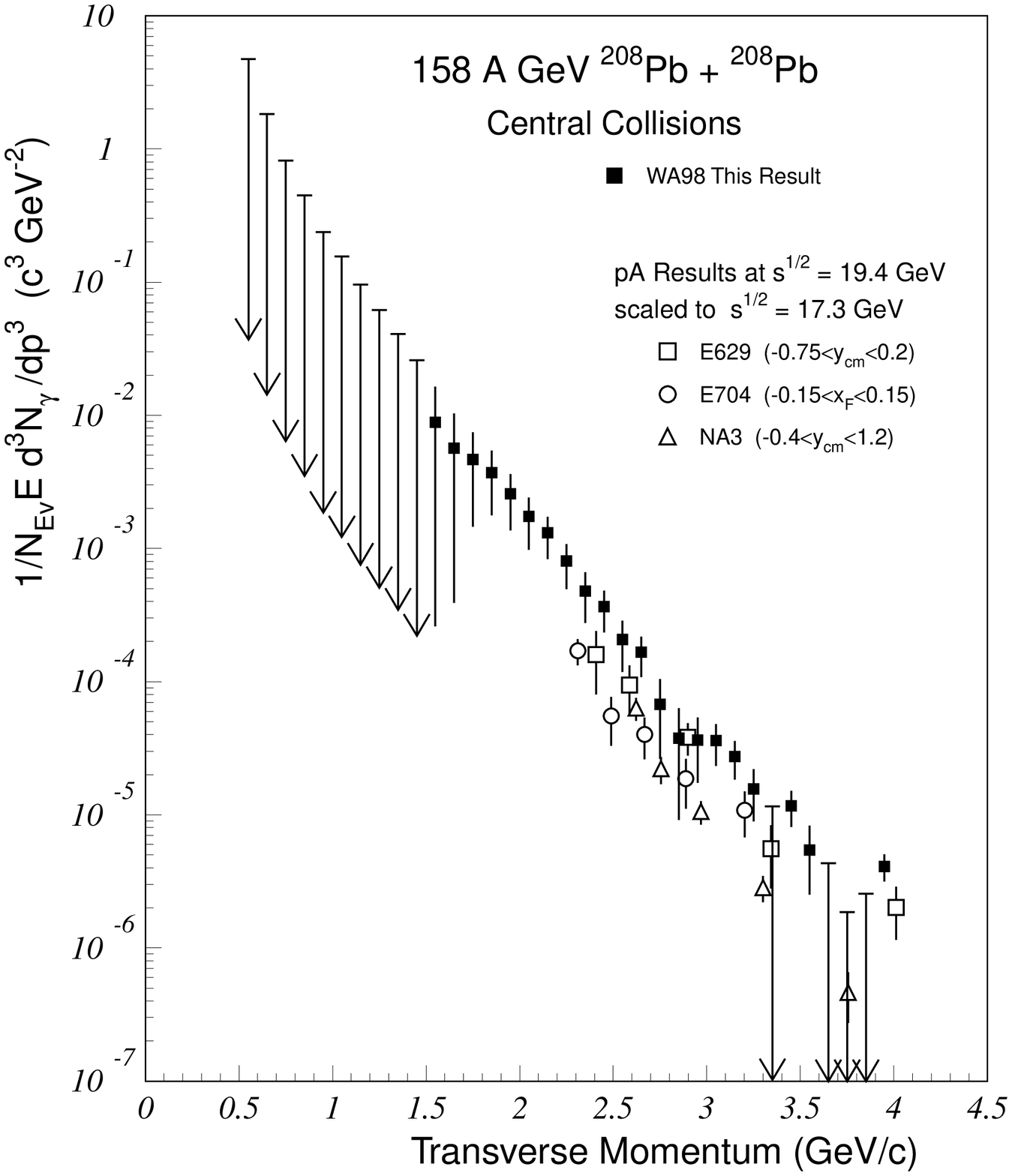}
\vspace{0.3cm}
\caption{The invariant direct photon multiplicity 
  for central 158~{\it A}~GeV $^{208}$Pb\/+\/$^{208}$Pb collisions.
  The error bars indicate the combined statistical and systematical
  errors.  Data points with downward arrows indicate unbounded 90\%
  CL upper limits.  Results of several direct photon measurements for
  proton-induced reactions have been scaled to central
  $^{208}$Pb\/+\/$^{208}$Pb collisions for comparison. 
  }
\label{fig:gamma_excess_cs}
\end{center}
\end{figure}

The ratio of measured photons to calculated background photons is 
displayed in Fig.~\ref{fig:gamma_excess} as a function of transverse 
momentum. The upper plot shows the ratio for peripheral collisions 
which is seen to be compatible with one, i.e. no indication of a 
direct photon excess is observed. The lower plot shows the same ratio for 
central collisions. It rises from a value of $\approx 1$ at low 
$p_{T}$ to exhibit an excess of about 20\% at high $p_{T}$.

A careful study of possible systematical errors is crucial for
the direct photon analysis. The largest 
contributions are from the $\gamma$ and $\pi^0$ identification 
efficiencies and the uncertainties related to the $\eta$ measurement. 
It should be emphasized
that the inclusive photon and neutral meson
(the basis for the background calculation) yields have been extracted from 
the same detector for exactly the same data sample. 
This decreases the sensitivity to many detector related errors and 
eliminates all errors associated  
with trigger bias or absolute yield normalization. 
Full details on 
the systematical error estimates are given in \cite{wa98photonslong}. 
The total $p_{T}$-dependent systematical errors are shown by the shaded 
regions in Fig.~\ref{fig:gamma_excess}. A significant photon 
excess is clearly observed in central collisions for $p_{T} > 1.5 \, 
\mathrm{GeV}/c$.

The final  invariant direct photon yield per central 
collision is presented in Fig.~\ref{fig:gamma_excess_cs}. 
The statistical and asymmetric systematical
errors of Fig.~\ref{fig:gamma_excess} are added in quadrature to
obtain the total upper and lower errors shown in
Fig.~\ref{fig:gamma_excess_cs}. An additional $p_T$-dependent error is
included to account for that portion of the uncertainty in the energy scale
which cancels in the ratios. In the case that the
lower error is less than zero a downward arrow is shown with the tail
of the arrow indicating the 90\% confidence level upper limit
($\gamma_{Excess}+1.28\,\sigma_{Upper}$).

No published prompt photon results exist for proton-induced reactions
at the $\sqrt{s}$ of the present measurement.
Instead, prompt photon yields for proton-induced reactions on fixed
targets at 200 GeV are shown in Fig.~\ref{fig:gamma_excess_cs} for
comparison \cite{plb:ada95,prl:mcl83,zpc:bad86}.
These results have been scaled for comparison with the present measurements 
according to the calculated average number 
of nucleon-nucleon collisions (660) for the central Pb+Pb event selection
and according to the beam energy under the assumption that $E
d^3\sigma_{\gamma}/dp^3 = f(x_T)/s^2$, where  
$x_T=2p_T/\sqrt{s}$ \cite{rmp:owe87}.  
This comparison indicates that the observed direct photon
production in central $^{208}$Pb\/+\/$^{208}$Pb collisions
has a shape similar to that expected for proton-induced reactions
at the same $\sqrt{s}$ but a yield which is enhanced.

\section{Summary}

The first observation of direct photons 
in ultrarelativistic heavy-ion collisions has been presented. 
While peripheral Pb+Pb 
collisions exhibit no significant photon excess, the 10\% most central 
reactions show a clear excess of direct photons in the range of 
$p_T$ greater than about $1.5 \, \mathrm{GeV}/c$. The invariant
direct photon multiplicity as a function of transverse momentum
was presented for central $^{208}$Pb\/+\/$^{208}$Pb 
collisions and compared to proton-induced results at similar
incident energy. The comparison indicates excess direct photon
production in central $^{208}$Pb\/+\/$^{208}$Pb collisions
beyond that expected from proton-induced reactions.  
The result suggests 
modification of the prompt photon production
in nucleus-nucleus collisions, or additional contributions from
pre-equilibrium or thermal photon emission.
The result should
provide a stringent test for different reaction scenarios,
including those with quark gluon plasma formation, and may provide  
information on the initial temperature attained in these collisions.

\end{document}